\documentclass[aps, prd, twocolumn, showpacs, floatfix, letterpaper, nofootinbib, superscriptaddress,longbibliography]{revtex4-2}

\usepackage{graphicx}
\usepackage{bm}
\usepackage{bbm}
\usepackage{amsfonts}
\usepackage{color}
\usepackage[T1]{fontenc}
\usepackage[latin1]{inputenc}
\usepackage{graphicx}
\usepackage[english]{babel}
\usepackage{amsmath}
\usepackage{amssymb}
\usepackage{makecell}
\usepackage{hyperref}
\usepackage{mathtools}
\usepackage{soul}
\usepackage[normalem]{ulem}
\usepackage{relsize}
\usepackage{orcidlink}
\usepackage[inline]{rotating}

\definecolor{green}{rgb}{0.19,0.64,0.54}
\definecolor{blue}{rgb}{0,0,1} \definecolor{reddish}{rgb}{0.65, 0.2,
  0.2} \definecolor{darkgreen}{rgb}{0.2,0.7,0.3}
\definecolor{darkblue}{rgb}{0.3,0.40,0.48}
\definecolor{gray}{rgb}{.8,.8,.8}

\hypersetup{,
  pdftitle={Time},
  pdfauthor={H. Bergeron, J.P Gazeau, P Malkiewicz and P Peter},
  colorlinks=true, 
  linkcolor=darkblue, 
  citecolor=darkgreen, 
  filecolor=reddish, 
  linktocpage=true
  }

\newcommand{\dd}{\mathrm{d}}

\newcommand{\ud}{\mathrm{d}}

\newcommand{\ba}{\begin{eqnarray}}
\newcommand{\ea}{\end{eqnarray}}

\newcommand{\ii}{\mathrm{i}}
\newcommand{\eqdef}{\stackrel{\mbox{\tiny def}}{=}}  
\newcommand{\ket}[1]{|\kern.3ex#1\kern.3ex\rangle}
\newcommand{\ketn}[1]{\kern.3ex#1\kern.3ex\rangle}
\newcommand{\bra}[1]{\langle\kern.3ex #1 \kern.3ex|}
\newcommand{\scalar}[2]{\langle\kern.3ex #1 \kern.3ex|\kern.3ex#2\kern.3ex\rangle}
\newcommand{\norm}[1]{\|\kern.3ex#1\kern.3ex \|}

\begin{document}

\title{A new class of exact coherent states: enhanced quantization of motion on the half-line.}

\author{Herv\'e Bergeron\orcidlink{0000-0002-6122-1953}} %
\email{herve.bergeron@universite-paris-saclay.fr } 
\affiliation{Institut des Sciences Mol\'{e}culaires d'Orsay (ISMO), UMR 8214 CNRS, Universit\'{e} Paris-Saclay, 91405 Orsay Cedex, France}

\author{Jean-Pierre Gazeau} \email{gazeau@apc.univ-paris7.fr}
\affiliation{APC, UMR 7164 CNRS, Universit\'{e} Paris Diderot, Sorbonne Paris Cit\'{e}, 75205 Paris, France}

\author{Przemys{\l}aw Ma{\l}kiewicz}
\email{Przemyslaw.Malkiewicz@ncbj.gov.pl}

\affiliation{National Centre for Nuclear Research, Pasteura 7, 02-093
  Warszawa, Poland}

\author{Patrick Peter\orcidlink{0000-0002-7136-8326}}%
\email{patrick.peter@iap.fr}

\affiliation{${\cal G}\mathbb{R}\varepsilon\mathbb{C}{\cal O}$ --
  Institut d'Astrophysique de Paris, CNRS \& Sorbonne Universit\'e,
  UMR 7095 98 bis boulevard Arago, 75014 Paris, France}

\begin{abstract}
We have discovered a class of dynamically stable coherent states for motion on the half-line. The regularization of the half-line boundary and the consequent quantum motion are expounded within the framework of covariant affine quantization, although alternative approaches are also feasible. The former approach is rooted in affine coherent states and offers a consistent semiclassical representation of quantum motion. However, this method has been known to possess two shortcomings: (a) the dependence of affine coherent states on the choice of a vector, denoted as 'fiducial vector' (which remains unspecified), introduces significant arbitrariness in boundary regularization, and (b) regardless of the choice of 'fiducial vector,' affine coherent states fail to evolve parametrically under the Schr\"odinger equation, thus limiting the accuracy of the semiclassical description. This limitation, in particular, hampers their suitability for approximating the evolution of compound observables. We demonstrate that a distinct and more refined definition of affine coherent states can simultaneously address both of these issues. In other words, these new affine coherent states exhibit parametric evolution only when the 'fiducial vector,' denoted as $\ket{\psi_0}$, possesses a highly specific character, such as being an eigenstate of a well-defined Hamiltonian. Our discovery holds significant relevance in the field of quantum cosmology, particularly in scenarios where the positive variable is the scale factor of the universe, and its regularized motion plays a crucial role in avoiding the big-bang singularity.
\end{abstract}

\maketitle

\section{Introduction}

This paper deals with the problem of quantization of a system on the half-line and of finding its exact quantum motion. Our motivation comes from quantum cosmology, where the scale factor of the universe is a positive variable and its regularized motion represents the avoidance of the big-bang singularity (see e.g. \cite{PhysRevD.108.043534}). We derive a new class of coherent states (defined on the half-line) that evolve parametrically under the Schr\"odinger equation. By the latter we mean that a complete set of solutions to the Schr\"odinger equation are expressible in terms of trajectories in the space of coherent states' parameters, and these trajectories satisfy simple Hamilton's equations generated by a ``semi-classical'' Hamiltonian. Unlike the standard coherent states (defined on the real line), the new coherent states do not solve the classical equations of motion but semi-classical ones that include an important quantum correction regularizing the behavior of the system at the boundary. Because of this regularizing effect of quantization and of the key role of the semi-classical Hamiltonian, we work in the framework of covariant affine quantization. We use the latter to derive and discuss our results, which can however be viewed independently of this particular framework, having applications beyond it.

Covariant affine quantization is a special case of a generic approach to quantization named \textit{Covariant Integral Quantization} (CIQ) \cite{berg2014, jpg2014, jpg2016,  jpg2015}. Introduced ten years ago, CIQ is based on operator-valued measures, with \textit{covariant} meaning that the quantization map intertwines classical (geometric operations) and quantum (unitary transformations) symmetries, while \textit{integral} refers to the fact that the map uses the resources of integral calculus, in order to implement the method for singular situations. Furthermore, CIQ is not only a method to build a quantum model from a classical theory, but it is also a tool to build a semi-classical portrait of quantum behavior.

CIQ includes the so-called Berezin-Klauder-Toeplitz quantization, and more generally coherent state quantization \cite{jpg2014, berezin1974, berezin1975}. On the mathematical level, CIQ is part of important developments such as time-frequency and wavelet analyses (see, for instance, Ref.~\cite{werner84,feichwer98,luefskrett18,berge22} and references therein). A famous example is the covariant integral quantization of the plane (phase space of the motion on the real line) based on the Weyl-Heisenberg group (WH), like Weyl-Wigner \cite{weyl1927, grossman1976, daubechies1980, daubechies1983} and (standard) coherent states quantization \cite{berezin1974}. Many other quantizations follow this framework \cite{berg2014, berg2018}.

Another important example of CIQ concerns the half-plane $\Pi_+ = \{ (q,p) \, | \, q >0, p \in \mathbb{R} \}$ viewed as the phase space for the motion on the half-line \cite{berg2014, jpg2016, berg2018, berg2020} (see also \cite{paul84} and references therein). The Lie group involved is that of affine transformations 
$$
x \mapsto (q,p) \cdot x := \frac{x}{q} + p, \quad q>0,
$$
of the real line. If $\Pi_+$ is equipped with the combination law $(q,p)\cdot (q', p') = \left(q q', p + p'/q \right)$, it can be identified with the affine group $\text{Aff}_+(\mathbb{R})$, the left invariant measure being $\dd \mu(q,p) = \dd q \dd p$.

In many previous papers involving the half-plane and devoted to quantum cosmology \cite{berg2014a, berg2015, berg2015a, berg2015b, berg2016, berg2020}, CIQ was implemented using Affine Coherent States (ACS) built from a Unitary Irreducible Representation (UIR) of the group $\text{Aff}_+(\mathbb{R})$ acting on the Hilbert space $\mathcal{H} = L^2(\mathbb{R}_+, \dd x)$ as $\hat{U}_{q,p}: \psi \in \mathcal{H} \mapsto \hat{U}_{q,p} \psi (x) = q^{-1/2} e^{\ii p x} \psi(x/q)$, 
with $\mathcal{H}$ the Hilbert space of the quantum representation (we work in units with $\hbar=c=1$). The ACS are defined as 
$$
\ket{q,p}_{\psi_0} \eqdef \hat{U}_{q,p} \ket{\psi_0} \in \mathcal{H},
$$
where $\ket{\psi_0} \in \mathcal{H}$ is a fixed normalized vector named the \textit{fiducial vector}. Besides the supplementary condition $\psi_0 \in L^2(\mathbb{R}_+, \dd x/x)$, the state $\ket{\psi_0}$ is a free parameter of the quantization procedure. Then the quantization map $f(q,p) \mapsto \hat{A}_f$ is implemented through the family of ACS $\{ \ket{q,p}_{\psi_0} \}_{(q,p) \in \Pi_+}$ \cite{berg2014, almeida2018}, the resulting operators $\hat{A}_f$ acting on the Hilbert space $\mathcal{H}$. The physical meaning of the fiducial vector becomes clear upon considering the quantization of the classical states represented by Dirac's delta functions on the phase space:
$$\delta(q_0,p_0)\mapsto \ket{q_0,p_0}\bra{q_0,p_0}_{\psi_0}.$$
They are thus replaced with the respective projector operators on $\mathcal{H}$ that depend on the fiducial vector. At the semi-classical level these sharply defined classical states become smooth probability distributions:
$$\delta(q-q_0,p-p_0)\mapsto |\bra{q_0,p_0}\ketn{q,p}_{\psi_0}|^2,$$
which again are determined by $\ket{\psi_0}$. These smooth probability distributions (e.g., Gaussians) encode the basic uncertainty and are used to regularize all the observables $f(q,p) \mapsto \check{f}(q,p)=\int\dd q_0\dd p_0~f(q_0,p_0) |\bra{q_0,p_0}\ketn{q,p}|^2$.

In order to recover the simple correspondence $\hat{A}_q = \hat{x}$ and
$\hat{A}_p = \hat{p}$, where $\hat{x} \psi(x) = x \psi(x)$ and $\hat{p}
\psi(x) = -\ii \psi'(x)$, automatically ensuring the commutation relation
$[ \hat{A}_q, \hat{A}_p]=\ii$, it suffices to impose a few relations on
the expectation values of $\ket{\psi_0}$. Let us notice that although
$\hat{x}$ is self-adjoint on $\mathcal{H}$, $\hat{p}$ is only symmetric and does not possess any self-adjoint extension on $\mathcal{H}$ \cite{reedsimon}. We also obtain the canonical correspondence $\hat{A}_{qp} = \frac12 (\hat{x} \hat{p} + \hat{p} \hat{x}) = \hat{d}$, i.e. the self-adjoint generator of dilations on $\mathcal{H}$. But the canonical correspondence is broken for $p^2$ since $\hat{A}_{p^2} \ne \hat{p}^2$. We instead have $\hat{A}_{p^2} = \hat{p}^2 + C_{\psi_0} \hat{q}^{-2}$ where $C_{\psi_0} >0$ is a positive constant depending on the choice of the fiducial vector $\ket{\psi_0}$. 

The operator $\hat{p}^2$  often being part of the Hamiltonian for many quantum systems, it should be made be self-adjoint.  There exist different self-adjoint extensions of $\hat{p}^2$ on the half-line,  depending on the boundary condition at $x=0$ that is not a part of canonical rules. Therefore, $\hat{p}^2$ is not uniquely defined as a self-adjoint operator, so that, if $\hat{H} \propto \hat{p}^2$,  the unitary evolution is not uniquely specified. On the other hand, if $\hat{A}_{p^2} = \hat{p}^2 + C_{\psi_0} \hat{q}^{-2}$ with $C_{\psi_0} > \frac34$, then $\hat{H}$ has a unique self-adjoint extension \cite{reedsimon, gesztesy} on $\mathcal{H}$, so the Hamiltonian and the quantum evolution are completely specified by the quantization procedure. At the semi-classical level, for any quantum operator $\hat{\mathcal{O}}$
the mapping $\hat{\mathcal{O}} \mapsto \bra{q,p} \hat{\mathcal{O}} \ket{q,p}$ gives a semi-classical picture of the quantum observable  $\hat{\mathcal{O}} $. More details about the semi-classical expressions and probabilistic aspects can be found in \cite{berg2014, berg2020}. Let us note finally that the affine group and related coherent states were also used for the quantization of the half-plane in works by J. R. Klauder, although from a different point of view and with a definite fiducial state $\psi_0$ selected along an algebraic condition (see \cite{klauder1970, klauder2011, fanuel2013} with references therein).

An interesting application of the ACS quantization lies in quantum cosmology, i.e. models of a homogeneous universe that in the classical theory suffer from the big-bang/big-crunch singularity. The regularizing potential $\propto\hat{q}^{-2}$ of the quantized Hamiltonian naturally acts as an ``anti-gravity'' component that
eventually overtakes the attractive force of gravity, bringing the contracting universe to a halt and triggering re-expansion \cite{berg2014a, berg2015, berg2015a, berg2015b, berg2016, berg2020}. One then extends these models by adding perturbations to homogeneity. The use of affine coherent states allows to conveniently account for the coupling of these perturbations to the quantized background spacetime via expectation values of some background quantities and next to solve the full dynamical system \cite{PhysRevD.103.083529,PhysRevD.105.023522,martin2022ambiguous}. 

Unfortunately, the approach discussed above crucially depends on the choice of the family of coherent states and on how well they approximate the exact dynamics. This is particularly true when the coupling is through a compound variable. Some effort has been undertaken to better control the accuracy of this approach by allowing some dynamics in the fiducial vectors, thereby making it less rigid and thus more adjustable to a given dynamics \cite{PhysRevD.98.026030}; although there are indisputable positive points to be credited to ACS quantization on the half-plane, some aspects of the method still need clear-cut justifications.

The first aspect concerns the arbitrariness of the fiducial vector $\ket{\psi_0}$. Since only a small number of constraints on $\ket{\psi_0}$ are necessary to obtain the results mentioned above -- and they concern only some expectation values, i.e. integrals over the state, so  $\ket{\psi_0}$ remains largely unspecified. Depending on the point of view about ``what a quantization procedure must be'', this can be seen as an advantage or a weakness. On the one hand, the main quantum observables $\hat{A}_f$ given by the procedure are the same, up to a few number of constraints on $\ket{\psi_0}$: one can consider the arbitrariness of $\ket{\psi_0}$ as a necessary mathematical feature to be able to deal with the quantization of all functions $f(q,p)$, while in fact the most important quantum observables are not modified by a change of $\ket{\psi_0}$. From this point of view, the arbitrariness of $\ket{\psi_0}$ is seen as a degree of freedom that allows to generate different possible ``complete'' quantum frameworks that share a common subset of quantum observables, each of these frameworks being mathematically consistent. In principle, only experiments could be able to select the good/best one. On the other hand, if one considers that a quantization procedure must directly provide ``the complete right quantum theory'' with ``certainty'', or with a finite list of unknown parameters that can be fixed by experiments, then the arbitrariness of $\ket{\psi_0}$ is a drawback: one expects that the mathematical expression of $\psi_0(x)$ should be specified as a function dependent on some unknown parameters. This entails that something is missing.

The second point concerns the dynamical properties of the ACS $\ket{q,p}_{\psi_0}$. First, examining the  dynamical properties at the quantum level means that,  at the classical level, we view the half-plane not only as a geometric domain invariant under affine transformations, but  also as the phase space (equipped with its Poisson bracket) of some system whose evolution is ruled by some classical Hamiltonian $H$. Therefore, the classical structure is much richer than the simple affine symmetry. Let us focus in the remainder on the case where $H=p^2$. If the vectors $\ket{q,p}_{\psi_0}$ are distinguished as ``special'', because allowing a mapping between classical and quantum pictures through ACS quantization, we can try to impose that this ``classical to quantum'' mapping is not only valid at a given time (affine symmetry), but also during evolution with time (dynamical symmetry). This means that we can try to impose that, at least up to a time-dependent phase factor, the $\ket{q,p}_{\psi_0}$ evolve parametrically through the Schr\"odinger equation, and this for some specific choice of the fiducial vector $\ket{\psi_0}$. Said differently, we demand that there exists a phase space trajectory $(q_t, p_t)$ and a phase $\phi(t)$ such that $\ii \partial_t \left[ e^{-\ii \phi(t)} \ket{q_t, p_t}_{\psi_0}\right] = e^{-\ii \phi(t)} \hat{H} \ket{q_t, p_t}_{\psi_0}$ for some specific choice of $\ket{\psi_0}$, where the quantum Hamiltonian $\hat{H}$ would be precisely the operator obtained from the ACS quantization, i.e. $\hat{H} = \hat{A}_{p^2}$. If possible, this enhanced framework would be a consistent way to fix the existing unease with the procedure. However, and this could be viewed as a drawback of the existing framework,  the ACS used in all papers cited above are not evolving parametrically whatever the choice of the fiducial vector $\ket{\psi_0}$, therefore this idea cannot be taken any further without modifying the ACS.

The purpose of this article is to prove that it is possible to define  the ACS quantization for a specific type of classical Hamiltonian and for a specific choice of the fiducial vector $\ket{\psi_0}$ evolving parametrically through the Schr\"odinger equation in such a way that it always gives the same basic quantum operators. One is left with a single unknown parameter in this procedure, namely the coefficient $C >0$ that appears in the repulsive term of $\hat{A}_{p^2} = \hat{p}^2 + C \hat{q}^{-2}$. The key point is to use a degree of freedom (a parameter) that exists in the definition of the UIR of the affine group $\text{Aff}_+(\mathbb{R})$,  parameter which is usually considered irrelevant and chosen to vanish; it turns out a different choice allows for the new effects we are interested in.

The article is organized as follows. In Section \ref{SecDef}, we define the new UIR of the affine group needed for our calculations. Sec.~\ref{SecNew} is devoted to the application of the ACS quantization to recover the most important quantum observables and in Sec.\ref{SecProof},  we prove  that for a special choice of $\ket{\psi_0}$, we obtain ACS that evolves parametrically. A final section \ref{SecConc} presents some concluding remarks.

\section{New definition of the ACS}\label{SecDef}

The affine coherent state quantisation discussed in the introduction can be generalised through the use of a seemingly innocuous and irrelevant parameter. We first show why this parameter may label physically different representations,  and justify a specific choice for its value. 

\subsection{The new framework}

The usual UIR of the affine group are defined on the Hilbert space $\mathcal{H} = L^2(\mathbb{R}_+, \dd x)$ as
\begin{equation}
\hat{U}_{q,p}: \psi \in \mathcal{H} \mapsto \hat{U}_{q,p} \psi (x) = 
\frac{e^{\ii p x}}{\sqrt{q}} \psi\left( \frac{x}{q}\right),
\end{equation}
which can be cast into the operator form
\begin{equation}
\hat{U}_{q,p} = e^{\ii p \, \hat{x}} e^{-\ii (\ln q) \, \hat{d}} \,,
\end{equation}
where $\hat{d} = \frac12 (\hat{x} \hat{p} + \hat{p} \hat{x})$ is the generator of dilations and $\hat{x} \psi(x) = x \psi(x)$ and $\hat{p} \psi(x) = -\ii \psi'(x)$ as usual. The relation $[\hat{d}, \hat{x}] = -\ii \hat{x}$  characterizes the Lie algebra of the affine group.  

Equivalent representations can be defined on the Hilbert spaces $\mathcal{H}_\alpha = L^2(\mathbb{R}, x^{-\alpha} \dd x)$, a choice which  imposes  $\psi\in \mathcal{H}_\alpha$ to obey  $\vert\psi(x)\vert \sim x^\gamma$ as $x\to 0$ with $\gamma > (\alpha-1)/2$. The choice $\mathcal{H} \equiv \mathcal{H}_{\alpha = 0}$ has been made because of the apparent absence of effect of $\alpha \ne 0$.  In turns out that such effects do exist if we use this degree of freedom differently. Indeed, these possibilities have a representation on the initial Hilbert space $\mathcal{H} = L^2(\mathbb{R}_+, \dd x)$ by choosing as generators of the group not the pair $(\hat{x}, \hat{d})$ but the pair $(\hat{x}^\alpha, \hat{d}/\alpha)$ with $\alpha >0$. Indeed we have again $[\hat{d}/\alpha, \hat{x}^\alpha] = - i \hat{x}^\alpha$ which specifies the Lie algebra of the affine group. Therefore there are infinitely many (equivalent) ways to represent the affine group on $\mathcal{H}$ dependent on a free parameter $\alpha >0$. They read
\begin{equation}
\label{unitV}
\hat{U}_{q,p}^{(\alpha)} = e^{\ii p \, \hat{x}^\alpha} e^{-\ii (\ln q) \, \hat{d}/\alpha} \,.
\end{equation}
Note that using the change of variable $x=e^y$ the dilation $\hat{d}$ becomes $\hat p_y=-\ii\frac{\dd~}{\dd y}$, canonical rules are restored, and the  classical phase space is $\mathbb{R}^2$. Then the above transform of the canonical pair is nothing more than the plane dilation $y\mapsto \alpha y\, , \, p_y\mapsto p_y/\alpha$. Together with plane rotation and upper triangular matrix  action  $y\mapsto y+ \mathsf{t} p_y\,,\, p_y\mapsto p_y$, these three actions are the Iwasawa factors of $\mathrm{SP}(2,\mathbb R)\cong \mathrm{SL}(2,\mathbb R)$. This observation paves the way to the use of more possibilities in dealing with representations of affine symmetries. 

In the following we are specially interested in the case $\alpha = 2$. The reason will be explained below.
Because we wish to preserve the correspondence $q \leftrightarrow \hat{x}$ and therefore $q^2 \leftrightarrow \hat{x}^2$, we first make a change of parameters in $\hat{U}^{(2)}_{q,p}$, substituting $q \mapsto q^2$. Since the pair $(q,p)$ is a  canonical pair, we also change  p as $p \mapsto p/(2q)$, to preserve canonicity. We then obtain a new realisation of the affine group UIR that we call $\hat{V}_{q,p}$ acting on the same $\hat{H} = L^2(\mathbb{R}_+, \dd x)$ as
\begin{equation}
\label{UnV}
\hat{V}_{q,p} = e^{\ii \frac{p}{2q} \, \hat{x}^2} e^{-\ii (\ln q^2) \, \hat{d}/2} = e^{\ii \frac{p}{2q} \, \hat{x}^2} e^{-\ii (\ln q) \, \hat{d}} \,.
\end{equation}
With this new parametrization of the affine group, the previous law of the group $(q,p).(q',p') = (qq', p+p'/q)$ is of course modified and we now have $(q,p).(q',p') = (qq', q' p + p'/q)$. Nevertheless, the left invariant measure remains unchanged, i.e.,  $\dd \mu(q,p) = \dd q \dd p$, because our new parametrization results from a canonical transformation of the half-plane.

This UIR $\hat{V}_{qp}$ of $\text{Aff}_+(\mathbb{R})$ is  square integrable, akin to the former one, which implies similar functional properties: picking some unit-norm vector $\psi_0 \in L^2(\mathbb{R}_+, \dd x) \cap L^2(\mathbb{R}_+, \dd x/x^2)$, we define ACS as previously
\begin{equation}
\ket{q,p}_{\psi_0} = \hat{V}_{q,p} \, \ket{\psi_0} \,.
\end{equation}
In the remainder, to lighten notations as there should be no ambiguity, we remove the label $\psi_0$, i.e.,  $\ket{q,p}_{\psi_0}
\to\ket{q,p}$. Setting
\begin{equation}
c_\gamma(\psi) = \int_0^{+\infty} \frac{\dd x}{x^{\gamma+2}} | \psi(x)|^2 \,.
\end{equation}
the fiducial vector $\ket{\psi_0}$ is admissible and square integrable if the constant $c_0(\psi_0)$ is finite, and one gets
\begin{equation}
\label{resolid}
\int_{\Pi_+} \frac{\dd q \dd p}{2\pi c_0(\psi_0)} \, \ket{q,p} \bra{q,p} = \mathbbm{1}, 
\end{equation}
which is the expression of the resolution of the identity operator $\mathbbm{1}$ in $\mathcal{H}$.

\subsection{Choice rationale}
\label{rationale}

We extend the pure geometric symmetry group (i.e., the affine group) to a larger dynamical group at both classical and quantum levels. Classically, the affine Lie algebra (through Poisson brackets) is usually assumed to be generated by the pair $( q, d=qp )$, but adding the assumption that the classical Hamiltonian is $H=p^2$, one can obtain a  larger closed algebra $\mathfrak{A}$ with the generators $( H, p, q ,d,1 )$. Unfortunately, this algebra is no longer closed if the Hamiltonian is changed into 
\begin{equation}
\label{}
 \tilde{H} = p^2 + C q^{-2},  
\end{equation}
and since non-canonical quantum corrections precisely involve such a  repulsive term $C q^{-2}$, it is impossible to keep the algebra $\mathfrak{A}$: the structure of the algebra must be the same for both classical and quantum systems because, by assumption, the symmetry group we are seeking must act at both levels.

The pair $(q^2, d/2 )$ also generates a representation of the Lie algebra of the affine group, and the triplet $(H, q^2, d/2 )$ is the basis of a closed Lie algebra. Furthermore, if we change from $H$ to $\tilde{H}$, the algebra generated by $( \tilde{H}, q^2, d/2 )$ remains closed with the same structure coefficients whatever the repulsive potential $C q^{-2}$. With this choice, the affine symmetry appears as a part of the same dynamical symmetry group which acts both classically and quantum mechanically.
In addition, the Lie-Poisson algebra generated by $( \tilde{H}, q^2, d/2 )$ is just the well-known $\mathfrak{sp}(2)\cong \mathfrak{sl}(1,1)\cong\mathfrak{su}(1,1)$ Lie algebra (see for instance \cite{gazol20}). At the classical level, the usual canonical representation of $\mathfrak{su}(1,1)$ consisting of the three generators $( k_0, k_1, k_2 )$ verifying $\{k_0, k_1 \} =  k_2$, $\{k_0, k_2 \} = -k_1$, $\{k_1, k_2\} = -k_0$ is recovered with $k_0 = \frac12(\tilde{H}+q^2/4)$, $k_1 = \frac12 \left[\cos\omega(\tilde{H}-q^2/4)+\sin\omega\,d\right]$ and $k_2 = \frac12\left[-\sin\omega(\tilde{H}-q^2/4)+\cos\omega\,d\right]$, where $\omega$ is an arbitrary angle.\\
The canonical generators of a $\mathrm{SU}(1,1)$ unitary representation on some Hilbert space are three self-adjoint operators $( \hat{K}_0, \hat{K}_1, \hat{K}_2 )$ verifying the commutation rules
\begin{equation}
[\hat{K}_0,\hat{K}_1]= \mathrm{i}\hat{K}_2\, , \ [\hat{K}_0,\hat{K}_2]= -\ii\hat{K}_1\,, \ [\hat{K}_1,\hat{K}_2]= -\mathrm{i}\hat{K}_0\,. 
\end{equation}
The Casimir operator of the representation is given by $\hat{\mathcal{Q}} = \hat{K}_1^2 +\hat{K}_2^2-\hat{K}_0^2$. It can be shown (see appendix \ref{append2}) that, in our case, $\hat{\mathcal{Q}} = \lambda \mathbbm{1}$, where the $c$-number $\lambda$ is directly related to the coefficient $C$ involved in the quantization of $p^2$, i.e. that leading to $\hat{A}_{p^2} = \hat{p}^2+C \hat{q}^{-2}$. In other words, the appearance of a repulsive term $C \ne0$ in the Hamiltonian corresponds to a change of UIR of  $\mathfrak{su}(1,1)$. This explains the key role of $\mathfrak{su}(1,1)$ in our problem. Furthermore, if we introduce the ladder operators 
\begin{equation}
\hat{K}_{\pm}= \hat{K}_2\mp \ii \hat{K}_1\,,\quad [\hat{K}_+,\hat{K}_-]= -2\hat{K}_0\, , 
\end{equation}
the operators $\hat{V}_{q,p}$ of \eqref{UnV} can be expressed in terms of $\hat{K}_{\pm,0}$ (see appendix \ref{append2})
 \begin{equation}
 \hat{V}_{q,p} = e^{(\xi \hat{K}_+ -\bar\xi \hat{K}_-)} \,e^{\mathrm{i}\theta \hat{K}_0}=e^{\ii\theta^{\prime} \hat{K}_0} e^{(\xi^{\prime} \hat{K}_+ -\bar{\xi^{\prime}} \hat{K}_-)}\, , 
  \end{equation}
where the coefficients $\xi$ and $\theta$ depend on $q$ and $p$. The  unitary operator $e^{(\xi \hat{K}_+ -\bar\xi \hat{K}_-)}$ is the SU$(1,1)$ analogous of the 
displacement operator in the Weyl-Heisenberg symmetry case, and was used by Perelomov to build his  SU$(1,1)$ coherent states \cite{perelomov86,perelomov:1972cmp}.

\section{New ACS quantization of the half-plane}\label{SecNew}

Having settled the framework, we now move to the actual quantization and the establishment of the semi-classical setup through which we can define meaningful trajectories.

\subsection{The quantized observables}
\label{secquantization}

From the resolution of the identity the covariant integral quantization follows \cite{berg2014, berg2018} for any function $f(q,p)$ as
\begin{equation}
f \mapsto \hat{A}_f = \int_{\Pi_+} \frac{\dd q \dd p}{2\pi c_0(\psi_0)} \, f(q,p) \,\ket{q,p} \bra{q,p} \,.
\end{equation}
If we assume the fiducial vector $\psi_0(x)$ to be a real function and rapidly decreasing on $\mathbb{R}^+$, i.e. $\psi_0(x) \in \mathbb{R}$ is $C^\infty$ on $\mathbb{R}_+$ and $\forall n,m \in \mathbb{N}$, $\lim_{x \to 0^+} x^{-n} \psi_0^{(m)}(x) = \lim_{x \to +\infty} x^n \psi_0^{(m)} (x) =0$, the basic quantized observables can be obtained easily (and the coefficients $c_\gamma(\psi_0)$ of \eqref{resolid} are finite for all $\gamma$). Without this assumption of rapid decrease, calculations need more caution because of possible divergencies, or supplementary terms coming from integration by parts at different levels. Details can be found in the appendix \ref{append1}.

We obtain first
\begin{equation}
\forall \alpha \in \mathbb{R}, \quad \hat{A}_{q^\alpha} = \frac{c_{\alpha}(\psi_0)}{c_0(\psi_0)} \,\, \hat{x}^\alpha \quad \implies \quad \hat{A}_q = \frac{c_1(\psi_0)}{c_0(\psi_0)} \, \, \hat{x} \,,
\end{equation}
and we find also
\begin{equation}
\hat{A}_p = \frac{c_1(\psi_0)}{c_0(\psi_0)} \, \, \hat{p} \,.
\end{equation}
Therefore, it suffices to add the supplementary constraint on the fiducial vector $c_1(\psi_0) = c_0(\psi_0)$, which is obtained by a simple rescaling of $\psi_0(x)$, to recover the canonical rule $[\hat{A}_q , \hat{A}_p] = i$ with $\hat{A}_q = \hat{x}$ and $\hat{A}_p = \hat{p}$.

The quantization of the generator of dilations $d=qp$ yields
\begin{equation}
\hat{A}_{qp} = \frac{c_2(\psi_0)}{c_0(\psi_0)} \,\, \hat{d} \quad \text{with} \quad \hat{d} = \frac12 \left( \hat{x} \hat{p} + \hat{p} \hat{x}\right).
\end{equation}
We recover the expected quantum generator of dilations up to a renormalization factor $c_2/c_0$. Let us remark that the renormalization factor $c_2/ c_0$ cannot be removed if we have already imposed $c_1/c_0 = 1$. If we try to impose at the same time $c_2=c_1=c_0$ the unique solution is $\psi_0(x)^2 = \delta(x-1)$, $\delta(x)$ being the Dirac distribution,  which is not acceptable. If $c_1/c_0 = 1$ then we have necessarily $c_2/c_0 >1$.

Quantization of the classical  Hamiltonian $H = p^2$ gives
\begin{equation}
\hat{H} = \hat{A}_{p^2} = \frac{c_2(\psi_0)}{c_0(\psi_0)} 
\left\{ p^2 + \left[ \frac{K}{c_2(\psi_0)} - \frac{3}{2} \right] \frac{1}{\hat{x}^2}  \right\},
\end{equation}
with
\begin{equation}
K = \int_0^\infty \frac{\dd y}{y^2} \psi_0'(y)^2 \,.
\end{equation}
While not obvious at first sight, the constant in front of $\hat{x}^{-2}$ is positive since we have (keeping the assumption of rapid decrease for $\psi_0$)
\begin{equation}
K - \frac{3}{2} c_2(\psi_0) = \int_0^\infty \dd y \left[ y^2 \phi'(y)^2 + \frac{1}{2} \phi(y)^2 \right] \end{equation}
where
\begin{equation}
\phi(y) = y \psi_0(1/y) \,.
\end{equation}
Therefore we recover the main feature of the ACS quantization, namely the appearance of a repulsive potential in the quantum Hamiltonian. 

\subsection{The semi-classical framework}

The semi-classical framework is obtained by the mapping $\hat{\mathcal{O}} \mapsto \bra{q,p} \hat{\mathcal{O}} \ket{q,p}$ with always $\ket{q,p} = \hat{V}_{q,p} \ket{\psi_0}$. In what follows, although we do not assume $\psi_0$ to be the vector already used in the quantization procedure, we keep the same generic assumptions for $\psi_0$. We easily obtain
\begin{equation}
\label{expecq}
\forall \alpha \in \mathbb{R}, \,\, \bra{q,p} \hat{x}^\alpha \ket{q,p} = c_{-\alpha-2}(\psi_0) \, q^\alpha 
\end{equation}
which implies 
\begin{equation}
\bra{q,p} \hat{x} \ket{q,p} = c_{-3}(\psi_0) \, q \,,
\end{equation}
and
\begin{equation}
\label{expecp}
\bra{q,p} \hat{p} \ket{q,p} = c_{-3}(\psi_0) \, p \,.
\end{equation}
With a re-scaling of $\psi_0$, it is possible to impose $c_{-3}(\psi_0) = 1$ in order to obtain the expected relations $\bra{q,p} \hat{x} \ket{q,p} =q$ and $\bra{q,p} \hat{p} \ket{q,p} =p$. Indeed, if we define $\psi_{0,\lambda}(x) = \lambda^{-1/2} \psi_0(x/\lambda)$, $\psi_{0,\lambda}$ is always a unit-norm vector and $c_{-3}(\psi_{0, \lambda}) = \lambda \, c_{-3}(\psi_0)$. Then choosing $\lambda = c_{-3}(\psi_0)^{-1}$ we obtain $c_{-3}(\psi_{0, \lambda}) = 1$.\\

The generator of dilation $\hat{d}$ is obtained in a similar fashion, namely
\begin{equation}
\bra{q,p} \hat{d} \ket{q,p} = c_{-4}(\psi_0) \, q p \,,
\end{equation}
and finally for $\hat{p}^2$
\begin{equation}
\label{expecp2}
\bra{q,p} \hat{p}^2 \ket{q,p} = c_{-4}(\psi_0) \, p^2 + \frac{C}{q^2} \end{equation}
with
\begin{equation}
C = \int_0^\infty \psi_0'(x)^2 \, \dd x
\end{equation} 
a positive definite constant.

\section{Stability in time of ACS}\label{SecProof}

From Sec. \ref{secquantization} above, we know that the quantum Hamiltonian obtained from the covariant affine  quantization with these new ACS is, in fact, that provided by the old procedure, up to some renormalization factor. Therefore, the details of the ACS quantization are not so important, and we assume in this part that the quantized Hamiltonian $\hat{H}$ is just
\begin{equation}
\label{hamil}
\hat{H}_\nu =  \hat{p}^2 + \frac{\nu^2-\frac14}{\hat{x}^2} \,,
\end{equation}
where the coefficient of the repulsive potential has been written as $\nu^2-\frac14$ with $\nu > \frac12$ for later convenience. As indicated in the introduction, we are looking for a fiducial vector $\psi_0$ such that the ACS $\ket{q_t, p_t}_{\psi_0}$ evolves parametrically through the Schr\"{o}dinger equation, for some semi-classical trajectory $(q_t, p_t)$ and up to some global phase factor $e^{-\ii \phi(t)}$, i.e. 
\begin{equation}
\label{paramschro}
i \partial_t \left[ e^{-\ii \phi(t)} \ket{q_t, p_t}_{\psi_0} \right] = e^{-\ii \phi(t)} \hat{H}_\nu \ket{q_t, p_t}_{\psi_0} \,.
\end{equation}
Furthermore, since a simple rescaling on $\psi_0$ is able to give $c_{-3}(\psi_0) = 1$, we assume in what follows that this rescaling has been done. To find all solutions of our problem, we split the argument in two parts: the first  is devoted to necessary conditions, and the second part to sufficient ones.

\subsection{Necessary conditions on semi-classical trajectories $(q_t, p_t)$}

    To begin with, let us assume that there exists a choice for $\psi_0$ such that Eq.~\eqref{paramschro} holds true. We now show that this implies that the semi-classical trajectories $(q_t, p_t)$ are those generated by the semi-classical Hamiltonian $H_\textsc{sc}(q,p) = c_{-4}(\psi_0)^{-1} \bra{q,p} \hat{H}_\nu \ket{q,p}$.\\

Let us define $\ket{\psi(t)} = e^{-\ii \phi(t)} \ket{q_t, p_t}$ satisfying the Schr\"{o}dinger equation \eqref{paramschro}. From Ehrenfest equation we have
\begin{equation}
\frac{\dd}{\dd t} \bra{\psi(t)} \hat{\mathcal{O}} \ket{\psi(t)} = \ii \bra{\psi(t)} [\hat{H}_\nu,  \hat{\mathcal{O}} ] \ket{\psi(t)} \,.
\end{equation}
Due to the expression \eqref{hamil} of $\hat{H}_\nu$ we obtain
\begin{subequations}
\begin{align}
\frac{\dd}{\dd t} \bra{\psi(t)} \hat{x} \ket{\psi(t)} & = 2 \bra{\psi(t)} \hat{p} \ket{\psi(t)}, \\
\frac{\dd}{\dd t} \bra{\psi(t)} \hat{p} \ket{\psi(t)} & = 2 \left(\nu^2-\frac14\right) \bra{\psi(t)} \hat{x}^{-3} \ket{\psi(t)}, 
\end{align}
\end{subequations}
and finally
\begin{equation}
\frac{\dd}{\dd t} \bra{\psi(t)} \hat{H}_\nu \ket{\psi(t)} = 0 \,.
\end{equation}
Using Eqs.~\eqref{expecq}, \eqref{expecp} and \eqref{expecp2}, together with the assumption $c_{-3}(\psi_0) = 1$, we obtain
\begin{subequations}
\begin{align}
\frac{\dd q_t}{\dd t} &= 2 p_t, \\
\frac{\dd p_t}{\dd t} &= 2 \left(\nu^2-\frac14\right) \frac{c_1(\psi_0)}{q_t^3},
\end{align}
\label{hamilsc1}
\end{subequations}
and therefore
\begin{equation}
\label{hamilsc2} \frac{\dd}{\dd t}  \left[ c_{-4}(\psi_0) p_t^2 + 
\frac{\left( \nu^2-\frac14 \right) c_0(\psi_0) + C}{q_t^2} \right] = 0,
\end{equation}
where $C$ was defined in \eqref{expecp2}. Eq.~\eqref{hamilsc1} imply that the possible trajectories $(q_t, p_t)$ must be those generated by the semi-classical Hamiltonian $H_\textsc{sc}(q,p)$, defined as
\begin{equation}
\label{schamilfromcs}
H_\textsc{sc}(q,p) = p^2 + \frac{\left( \nu^2-\frac14 \right) c_1(\psi_0)}{q^2},
\end{equation}
up to an arbitrary and irrelevant additive constant. Consistency of the above with Eq.~\eqref{hamilsc2} implies that the constraint on $\psi_0$
\begin{equation}
\begin{split}
C & = \int_0^\infty \psi_0'(x)^2 \dd x \\
& = \left( \nu^2-\frac14 \right)
\left[ c_1(\psi_0) c_{-4}(\psi_0) - c_0(\psi_0) \right]
\end{split}
\label{psi0constr}
\end{equation}
must hold. If this constraint is not fulfilled, equations \eqref{hamilsc1} and \eqref{hamilsc2} are incompatible and therefore the ACS cannot evolve parametrically with the Schr\"{o}dinger equation. If the constraint \eqref{psi0constr} is fulfilled, then it may be possible to have ACS evolving parametrically with the Schr\"{o}dinger equation, and in that case the trajectories are necessarily given by the Hamiltonian of \eqref{schamilfromcs} which is just $H_\textsc{sc}(q,p) = c_{-4}(\psi_0)^{-1} \bra{q,p} \hat{H}_\nu \ket{q,p}$.

\subsection{Finding $\psi_0$}

Let us now consider in more details the conditions required for the existence of the fiducial state.

\subsubsection{Necessary Conditions}

We assume that some $\psi_0$ exists satisfying Eq.~\eqref{paramschro} with the rescaling $c_{-3}(\psi_0) = 1$. We then have
\begin{equation}
\label{eqschro1}
 i \, \hat{V}_{q_t p_t}^\dagger \partial_t \left[ e^{-\ii \phi(t)} \hat{V}_{q_t p_t} \right] \ket{\psi_0} = e^{-\ii \phi(t)} \hat{V}_{q_t p_t}^\dagger  \hat{H}_\nu \hat{V}_{q_t p_t} \ket{\psi_0} \,.
\end{equation}
Using the relations \eqref{hamilsc1} that necessarily hold true, and given the commutation relation between $\hat{x}^2$ and $\hat{d}$, we first deduce that
\begin{widetext}
\begin{equation}
\label{nccond1}
e^{\ii \phi(t)} \ii \, \hat{V}_{q_t p_t}^\dagger \partial_t \left[ e^{-\ii \phi(t)} \hat{V}_{q_t p_t} \right] = \phi'(t) + p_t^2 \hat{x}^2  - \frac{\left( \nu^2-\frac14 \right) c_1(\psi_0)}{q_t^2} \, \hat{x}^2 + \frac{2 p_t}{q_t} \hat{d}.
\end{equation}
Besides, we find
\begin{equation}
\label{nccond2}
\hat{V}_{q_t p_t}^\dagger  \hat{H}_\nu \hat{V}_{q_t p_t} = \frac{1}{q_t^2} \hat{p}^2 + p_t^2 \hat{x}^2 + \frac{2 p_t}{q_t} \hat{d} + \frac{\nu^2-\frac14}{q_t^2 \, \hat{x}^2},
\end{equation}
so that Eq.~\eqref{eqschro1} becomes
\begin{equation}
\label{necesseq1}
\left[ \hat{p}^2 + \frac{\nu^2-\frac14}{\hat{x}^2} + \left( \nu^2-\frac14 \right) c_1(\psi_0) \, \hat{x}^2 \right] \ket{\psi_0} = q_t^2 \phi'(t) \ket{\psi_0} \,.
\end{equation}
This implies that $q_t^2 \phi'(t) = \omega$ where $\omega$ is an eigenvalue of a fixed operator $\hat{H}_0$ which is nothing but the radial Hamiltonian of a 3D harmonic oscillator, and furthermore, $\psi_0$ is the eigenvector of $\hat{H}_0$ associated to $\omega$. Let us remark that in this case, $\psi_0(x)$ is not a rapidly decreasing function on $\mathbb{R}_+$ because $\psi_0(x) \propto x^{\alpha}$ when $x \to 0$. Nonetheless, because our initial assumption of rapid decrease was just a way to simplify proofs and was not in fact mandatory, this choice of $\psi_0$ turns out to be  completely valid: the different constants $c_\gamma(\psi_0)$ are merely only defined for some domain of $\gamma$ bounded above.

Taking into account the equations of motion \eqref{hamilsc1} the condition $q_t^2 \phi'(t) = \omega$ yields
\begin{equation}
\label{phisol}
q_t^2 \phi'(t) = \omega \quad \implies \quad \phi(t) = \frac{\omega}{2\sqrt{\left( \nu^2-\frac14 \right) c_1(\psi_0)}} \arctan  \left[ \frac{q_t p_t}{\sqrt{\left( \nu^2-\frac14 \right) c_1(\psi_0)}} \right],
\end{equation}
in which we used the relation $\dfrac{\dd}{\dd t} (q_t p_t) = 2 H_\text{sc}(q_t, p_t)$ stemming from \eqref{hamilsc1} and \eqref{schamilfromcs}. Therefore, the phase factor $\phi(t)$ is itself a function of $q$ and $p$, namely
\begin{equation}
e^{- \ii \phi(t)} = \left\{ \exp \left[-2i \arctan \frac{q_t p_t}{\sqrt{\left( \nu^2-\frac14 \right) c_1(\psi_0)}} \right] \right\}^{\frac14 \omega \left[\left( \nu^2-\frac14 \right) c_1(\psi_0)\right]^{-1/2}} ,
\end{equation}
which can also be written as
\begin{equation}
e^{- \ii \phi(t)} = \left[ \frac{\sqrt{\left( \nu^2-\frac14 \right) c_1(\psi_0)}-\ii q_t p_t}{\sqrt{\left( \nu^2-\frac14 \right) c_1(\psi_0)} + \ii q_t p_t} \right]^{\frac14 \omega \left[\left( \nu^2-\frac14 \right) c_1(\psi_0)\right]^{-1/2}}.
\end{equation}
\end{widetext}

\subsubsection{Sufficient conditions}

Reciprocally, let us assume that $\hat{H}_0$ is the following Hamiltonian
\begin{equation}
\label{hamil3doscll}
\hat{H}_0 = \hat{p}^2 + \frac{\nu^2-\frac14}{\hat{x}^2} + \xi^2 \hat{x}^2, \quad \text{with} \quad \nu >\frac12 \ \hbox{and} \ \xi >0,
\end{equation}
and let $\ket{\psi_0}$ be a normalized eigenvector of $\hat{H}_0$ with eigenvalue $\omega$, i.e.
\begin{equation}
\label{eigeneq}
\hat{H}_0 \ket{\psi_0}  = \omega \ket{\psi_0} \,.
\end{equation}
We know from the general properties of eigenfunctions that $\psi_0(x)$ are all real as required.

In Eq.~\eqref{eigeneq}, $\xi$ determines the length scale of the problem. If we now impose $c_{-3}(\psi_0) = 1$ as before, this constraint defines a relation between $\xi$ and $\nu$ that uniquely specifies $\xi$ as a function of $\nu$: we call name $\xi_\nu$ this specific value set $\xi = \xi_\nu$ from now on. The constraint $c_{-3}(\psi_0) = 1$ is thus automatically fulfilled.

Provided the relation
\begin{equation}
\label{firsteq}
C + \left( \nu^2 - \frac14 \right) c_0(\psi_0) + \xi_\nu^2 c_{-4}(\psi_0) = \omega
\end{equation}
holds, in which
\begin{equation}
\quad C = \int_0^\infty \psi'_0(x)^2 \dd x,
\end{equation}
we have that $\bra{\psi_0} \hat{H}_0 \ket{\psi_0} = \omega $. Besides, since $\ket{\psi_0}$ is an eigenvector of $\hat{H}_0$, we know that for any observable $\hat{\mathcal{O}}$, we have $\bra{\psi_0} [\hat{H}_0, \hat{\mathcal{O}} ] \ket{\psi_0} = 0$. Applying this relation for $\hat{\mathcal{O}} = \hat{p}$ and $\hat{\mathcal{O}} = \hat{d}$, we obtain two new relations
\\

\begin{subequations}
\begin{align}
\label{secondeq0}  \left( \nu^2 - \frac14 \right) c_1(\psi_0) -& \xi_\nu^2 c_{-3}(\psi_0) = 0,\\
\label{thirdeq} C+\left( \nu^2 - \frac14 \right) c_0(\psi_0) - &\xi_\nu^2 c_{-4}(\psi_0) = 0.
\end{align}
\end{subequations}
Due to the scaling $c_{-3}(\psi_0) = 1$, Eq.~\eqref{secondeq0} yields
\begin{equation}
\label{secondeq}
 \left( \nu^2 - \frac14 \right) c_1(\psi_0) = \xi_\nu^2,
\end{equation}
while from \eqref{firsteq} and \eqref{thirdeq}, we obtain
\begin{equation}
c_{-4}(\psi_0) = \frac{\omega}{2 \xi_\nu^2} \,,
\end{equation}
as well as
\begin{align}
C = & \frac{\omega}{2} - \left(\nu^2 - \frac14\right) c_0(\psi_0) \nonumber
\\ = & \left(\nu^2 - \frac14\right) \left[
c_1(\psi_0) c_{-4}(\psi_0) - c_0(\psi_0)\right],
\end{align}
so that the (mandatory) constraint \eqref{psi0constr} is also fulfilled.

Let us assume that the semi-classical trajectory $(q_t, p_t)$ is obtained from the semi-classical Hamiltonian $H_\textsc{sc}(q,p)$, defined as
\begin{equation}
H_\textsc{sc}(q,p) = p^2 + \frac{\xi_\nu^2}{q^2} = p^2 + 
\left(\nu^2 - \frac14\right) \frac{ c_1(\psi_0)}{q^2} \,,
\end{equation}
the last equality stemming from \eqref{secondeq}. The Hamiltonian $H_\textsc{sc}(q,p)$ coincides with that of Eq.~\eqref{schamilfromcs} and therefore, the Hamilton equations derived from $H_\textsc{sc}(q,p)$ satisfy Eqs.~\eqref{hamilsc1}. Thus, Eqs.~\eqref{nccond1} and \eqref{nccond2} of the previous section  are valid, and therefore $\ket{\psi_0}$ solves our problem if (and only if) Eq.~\eqref{necesseq1} holds true, which is equivalent to $q_t^2 \phi'(t) = \omega$. Taking into account \eqref{phisol} and \eqref{secondeq}, we finally obtain $\phi(t)$ as
\begin{equation}
\phi(t) = \frac{\omega}{2 \xi_\nu} \arctan \left(  \frac{q_t p_t}{\xi_\nu} \right) \,,
\end{equation}
and
\begin{equation}
e^{-\ii \phi(t)} = \left( \frac{\xi_\nu - \ii q_t p_t}{\xi_\nu + \ii q_t p_t} \right)^{\frac{\omega}{4 \xi_\nu}} \,.
\end{equation}
The above lines of arguments show that the fiducial vectors generating affine coherent states stable in time are exactly all eigenvectors of $\hat{H}_0$. Since the eigensystem for $\hat{H}_0$ is completely solvable, we can go a step further with explicit formula.

\subsection{Summary}

\begin{figure}[t]
\begin{center}
\includegraphics[scale=0.205]{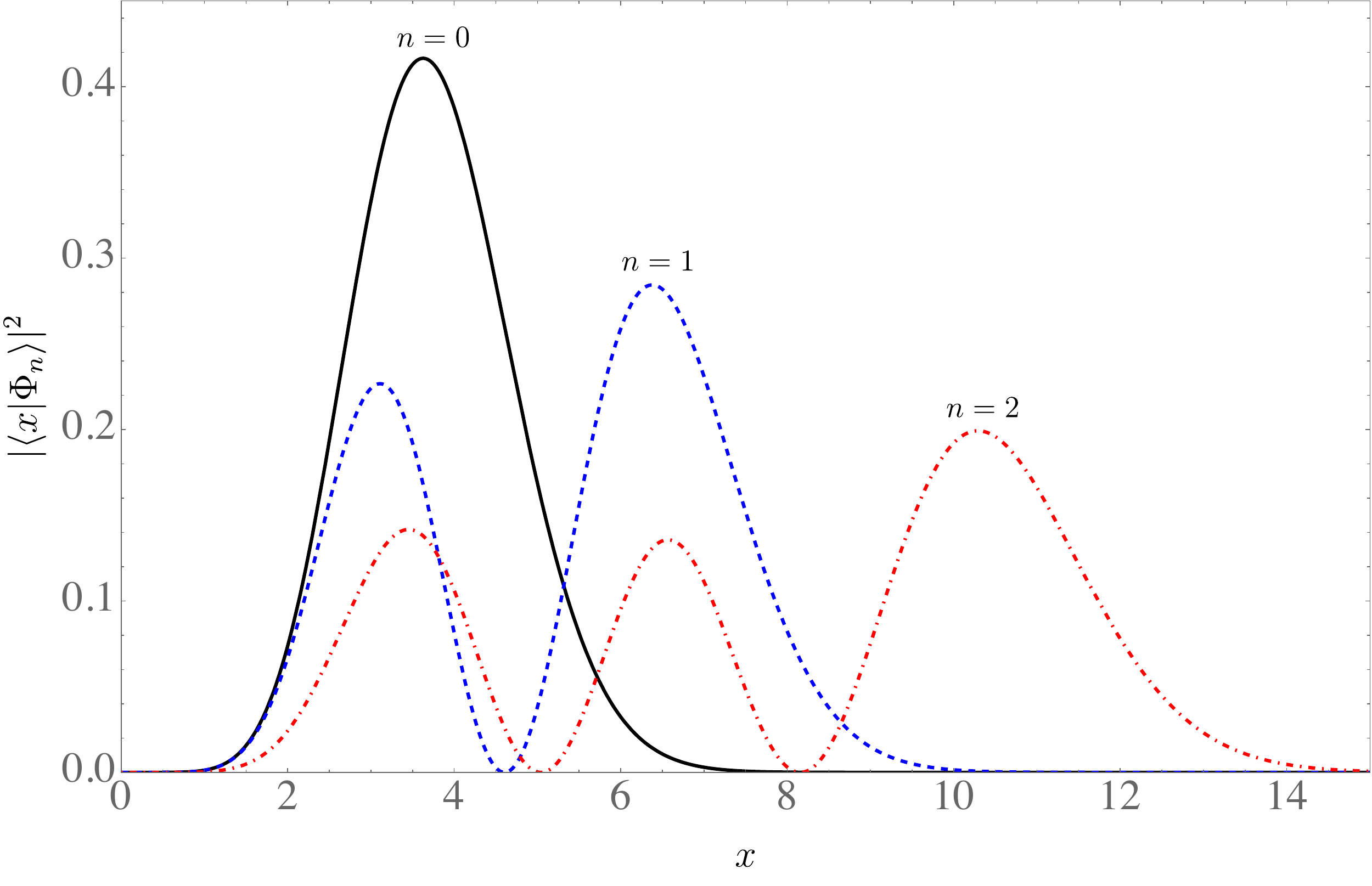} 
\caption{\label{fig3} The probability distributions of positions for the first three fiducial vectors: $|\langle x|\Phi_n\rangle |^2$ for $n=0$, $n=1$ and $n=2$ ($\nu=3$ and $H_\textsc{sc}=1$).}
\end{center}
\end{figure}

Since the solutions to our problem are given by the eigenvectors (and the eigenvalues) of the Hamiltonian~\eqref{hamil3doscll}, we begin by recalling some known results about it.

\subsubsection{Definitions}

The solutions of $\hat{H}_0 \ket{\Phi_n} = \omega_n \ket{\Phi_n}$ where $n \in \mathbb{N}$, and the $\ket{\Phi_n}$ being normalized, are
\begin{align}
\omega_n &= 2 \xi(2 n + \nu +1) , \\
\Phi_n(x) &= \sqrt{\frac{2 \,\, n!}{\Gamma(\nu+n+1)}} \,\, \xi^{\frac{\nu+1}{2}} \, x^{\nu+1/2} L_n^\nu\left( \xi x^2\right) e^{- \frac{1}{2} \xi x^2} \,,
\end{align}
where the $L_n^\nu(y)$ are Laguerre polynomials and we have defined $\Phi_n(x) \eqdef \langle x| \Phi_n \rangle$. The first three fiducial vectors are plotted in Fig. \ref{fig3}.

Let us introduce for convenience the functions $G_n(\alpha, \nu)$ for $\alpha + \nu >0$, $\nu >0$,  and $n \in \mathbb{N}$
\begin{equation}
G_n(\alpha, \nu) = \frac{n!}{\Gamma(\nu+n+1)} \int_0^\infty x^{\nu+\alpha-1} L_n^\nu(x)^2 e^{-x} \, \dd x \,.
\end{equation}
Although they can be easily computed for each value of $n$, there does not exist, to the best of our knowledge, a generic formula in terms of usual special functions. The normalization factor in front of the integral has been chosen such that $G_n(1, \nu) = 1$ and $G_0(\alpha, \nu) = \Gamma(\nu+\alpha)/\Gamma(\nu+1)$. With this definition we obtain
\begin{equation}
\label{cgamma}
c_\gamma(\Phi_n) = \xi^{1+\gamma/2} G_n\left( -\frac{\gamma}{2}, \nu\right),
\end{equation}
so that the scaling condition $c_{-3}(\Phi_n) = 1$ previously used gives, for $\xi_{\nu,n}$ 
\begin{equation}
\label{xinun}
\xi_{\nu,n} = \left[ G_n\left(\frac32, \nu\right)\right]^2.
\end{equation}
To end with the useful definitions, we call $\tilde{\omega}_{\nu,n}$ the values of $\omega_n$ for $\xi = \xi_{\nu,n}$, i.e.
\begin{equation}
\tilde{\omega}_{\nu,n} = 2 \xi_{\nu,n} (2n + \nu+1) = 2 \, \left[ G_n\left(\frac32, \nu\right)\right]^2 (2n + \nu+1).
\end{equation}
We are now in a position to summarize the results of the previous sections. 

\subsubsection{General case}

\begin{figure}[!ht]
\begin{center}
\includegraphics[scale=0.22]{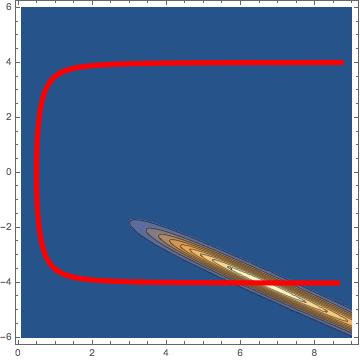} 
\includegraphics[scale=0.22]{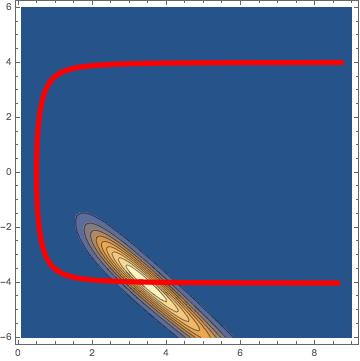}
\includegraphics[scale=0.22]{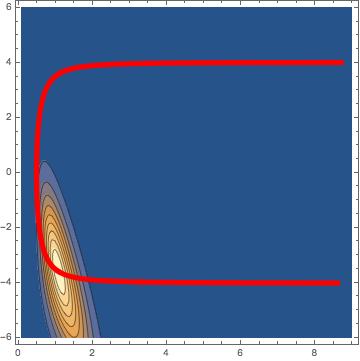}\\
\includegraphics[scale=0.22]{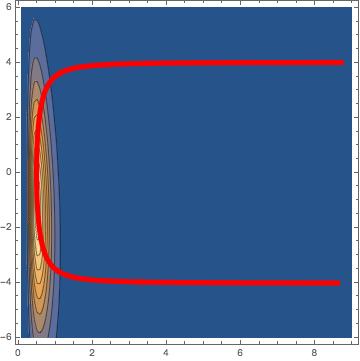}
\includegraphics[scale=0.22]{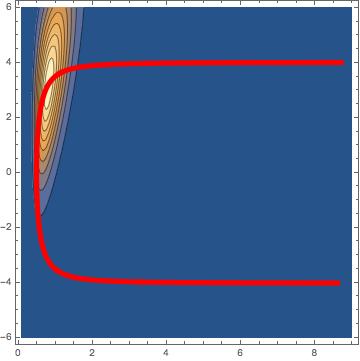}
\includegraphics[scale=0.22]{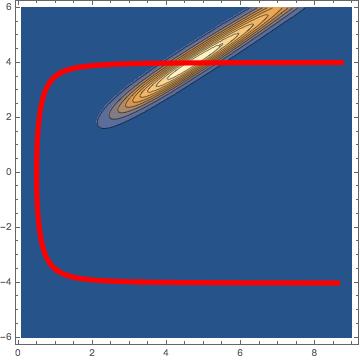}
\caption{\label{fig1} Some phase space portrait of $\rho_{\psi_t}(q,p)$ for $\nu=3$. Time $t$ is increasing from top left to bottom right. The semi-classical trajectory in red is going through the point $(q=5, p=-4)$.}
\end{center}
\end{figure}

Given the quantum Hamiltonian $\hat{H}_\nu$  \eqref{hamil}, all families of ACS $e^{-\ii \phi_{q,p}} \ket{q,p}_{\psi_0}$ evolving parametrically by the Schr\"{o}dinger equation as $\ii \partial_t \left( e^{-\ii \phi_{q_t, p_t}} \ket{q_t, p_t}_{\psi_0} \right) = e^{-\ii \phi_{q_t, p_t}} \hat{H}_\nu \ket{q_t, p_t}_{\psi_0} $ depend on an integer $n$ and of course on the parameter $\nu$ of $\hat{H}_\nu$. If we introduce the notation $\ket{q,p; \nu, n} \eqdef e^{-\ii \phi_{q, p}} \ket{q, p}_{\psi_0}$ that makes explicit all the parameters involved, we have the complete formula
\begin{widetext}
\begin{equation}
\scalar{x}{q,p;\nu, n} = \sqrt{\frac{2 \,\, n!}{\Gamma(\nu+n+1)}} \left( \frac{\xi_{\nu, n} - \ii q p}{\xi_{\nu, n} + \ii q p} \right)^{\frac12 \left(2n+\nu+1\right)} \xi_{\nu,n}^{\frac{\nu+1}{2}} \, \frac{x^{\nu+1/2}}{q^{\nu+1}} L_n^\nu\left( \xi_{\nu,n}  \frac{x^2}{q^2} \right) \exp \left[ - \frac{1}{2} (\xi_{\nu,n} - \ii \, q p) \frac{x^2}{q^2} \right]
\end{equation}
The semi-classical trajectories $(q_t, p_t)$ compatible with the Schr\"{o}dinger equation (with Hamiltonian $\hat{H}_\nu$) for a family $\{ \ket{qp; \nu, n} \}_{(q,p) \in \Pi_+}$ are those resulting from the following semi-classical Hamiltonien $H_\text{sc}^{(\nu,n)} (q,p)$ 
\begin{equation}
H_\text{sc}^{(\nu,n)}(q,p) = p^2 + \frac{\xi_{\nu,n}^2}{q^2} \,.
\end{equation}
Furthermore, we have
\begin{equation}
\bra{q, p; \nu, n} \hat{x} \ket{q, p; \nu, n} = q \quad \text{and} \quad \bra{q, p; \nu, n} \hat{p} \ket{q, p; \nu, n} = p \,,
\end{equation}
and, in addition
\begin{equation}
\bra{q, p; \nu, n} \hat{H}_\nu \ket{q, p; \nu, n} = c_{-4}(\Phi_n) H_\textsc{sc}^{(\nu,n)} (q,p) = \frac{\tilde{\omega}_{\nu,n}}{2 \xi_{\nu,n}^2} H_\textsc{sc}^{(\nu,n)} (q,p) = \frac{2n+\nu+1}{\xi_{\nu,n}}  H_\text{sc}^{(\nu,n)} (q,p)  \,.
\end{equation}
For each pair $(\nu,n)$, the family of states $\{\ket{q,p;\nu n} \}_{(q,p) \in \Pi_+}$ solves the identity according to the general formula \eqref{resolid}, i.e.
\begin{equation}
\int_{(q,p)\in \Pi_+} \frac{\dd q \dd p}{2 \pi c_0(\nu,n)} \, \ket{q,p; \nu, n} \bra{q,p; \nu, n} = \mathbbm{1}\,, \quad \text{with} \quad c_0(\nu,n) = \left[ G_n\left(\frac32, \nu\right)\right]^2 G_n(0,\nu),
\end{equation}
the expression of $c_0(\nu,n)$ resulting from Eqs.~\eqref{cgamma} and \eqref{xinun}. Therefore, for any $\ket{\psi} \in \mathcal{H}$ we have
\begin{equation}
\ket{\psi} = \int_{\Pi_+} \frac{\dd q \dd p}{2 \pi c_0(\nu,n)} \, \psi(q,p) \, \ket{q,p; \nu,n} \quad \text{with} \quad \psi(q,p) = \scalar{q,p; \nu,n}{\psi} \,.
\end{equation}
It follows that the time-dependent vector $\ket{\psi(t)} = e^{-\ii \hat{H}_\nu \, t} \ket{\psi}$ verifies
\begin{equation}
\label{timdeppsi}
\ket{\psi(t)} = e^{-\ii \hat{H}_\nu \, t} \ket{\psi} = \int_{\Pi_+} \frac{\dd q \dd p}{2 \pi c_0(\nu,n)} \, \psi_t(q,p)\,  \ket{q,p; \nu,n} \quad \text{with} \quad \psi_t(q,p) = \bra{q,p; \nu,n} e^{-\ii \hat{H}_\nu \, t} \ket{\psi} \,.
\end{equation}
Let us define $(q,p) \mapsto \left[ Q_t(q,p), P_t(q,p) \right]$ the Hamiltonian flow resulting from the semi-classical Hamiltonian $H_\textsc{sc}^{(\nu,n)}(q,p)$. Since the states $\ket{q,p; \nu,n}$ evolve parametrically, we have $e^{-i \hat{H}_\nu \, t} \, \ket{q,p; \nu, n} = \ket{Q_t(q,p), P_t(q,p); \nu, n}$ from which it follows that
\begin{equation}
\psi_t(q,p) = \bra{q,p; \nu,n} e^{-\ii \hat{H}_\nu \, t} \ket{\psi} = \scalar{Q_{-t}(q,p), P_{-t}(q,p); \nu, n}{\psi} = \psi\left[ Q_{-t}(q,p), P_{-t}(q,p)\right].
\end{equation}
Thus, the time-dependent vector $\ket{\psi(t)}$ of \eqref{timdeppsi}, written in the ACS (overcomplete) basis $\ket{q,p;\nu,n}$ possesses coefficients $\psi_t(q,p)$ of splitting that follow exactly the classical Liouville equation (but with a semi-classical Hamiltonian)
\begin{equation}
\frac{\partial \psi_t}{\partial t} = - \left\{\psi_t, H_\textsc{sc}^{(\nu,n)} \right\}.
\end{equation}
Therefore, the use of the ACS $\ket{q,p; \nu,n}$ allows (as desired) to completely intertwine quantum and classical evolution of states for the quantum Hamiltonian $\hat{H}_\nu$.

\subsubsection{Special case $n=0$}

In the particular case $n=0$, the previous generic formula can be simplified, giving
\begin{equation}
\scalar{x}{q,p;\nu, 0} = \sqrt{\frac{2}{\Gamma(\nu+1)}} \left( \xi_{\nu,0} \, \frac{\xi_{\nu,0} - \ii q p}{\xi_{\nu,0} + i q p} \right)^{\frac{\nu+1}{2}}\frac{x^{\nu+1/2}}{q^{\nu+1}} \exp \left[ - \frac{1}{2} (\xi_{\nu,0} - \ii q p) \frac{x^2}{q^2} \right] \quad \text{with} \quad \xi_{\nu,0} = \left[ \frac{\Gamma(\nu+3/2)}{\Gamma(\nu+1)} \right]^2 \,.
\end{equation}
The semi-classical trajectories $(q_t, p_t)$ compatible with the Schr\"{o}dinger equation are now coming from the semi-classical Hamiltonien $H_\text{sc}^{(\nu,0)}(q,p)$ 
\begin{equation}
H_\text{sc}^{(\nu,0)}(q,p) = p^2 + \frac{\xi_{\nu,0}^2}{q^2} \,.
\end{equation}
We keep the relations
\begin{equation}
\bra{q, p; \nu,0} \hat{x} \ket{q, p; \nu,0} = q \quad \text{and} \quad \bra{q, p; \nu,0} \hat{p} \ket{q, p; \nu,0} = p \,,
\end{equation}
and
\begin{equation}
\bra{q, p; \nu,0} \hat{H}_\nu \ket{q, p; \nu,0} = \frac{\nu+1}{\xi_{\nu,0}} H_\text{sc}^{(\nu,0)}(q,p) \,.
\end{equation}
Furthermore, since the vectors $\ket{q,p;\nu, 0}$ solve the identity, the function $\rho_\psi(q,p) = |\scalar{q,p; \nu,0}{\psi}|^2/[2\pi c_0(\nu,0)]$ is a semi-classical probability density that gives a phase-space portrait of any quantum state $\ket{\psi}$. In particular, we can obtain a phase space portrait for $\ket{\psi_t} = \ket{q_t,p_t; \nu, 0}$ at any time $t$ along a semi-classical trajectory. A straightforward calculation gives
\begin{equation}
|\scalar{q', p'; \nu,0}{q, p, \nu, 0}|^2 = \frac{(2 \xi_{\nu,0})^{2 \nu+2}}{\left[ \xi_{\nu,0}^2 \left( \frac{q'}{q}+\frac{q}{q'} \right)^2+ (q p'-q' p)^2 \right]^{\nu+1}}.
\end{equation}
An example of $\rho_{\psi_t}(q,p)$ for $\ket{\psi_t} = \ket{q_t,p_t; \nu, 0}$ is given in figure \ref{fig1}.
\end{widetext}

\subsubsection{Special case $n=1$}

\begin{figure}[!ht]
\begin{center}
\includegraphics[scale=0.33]{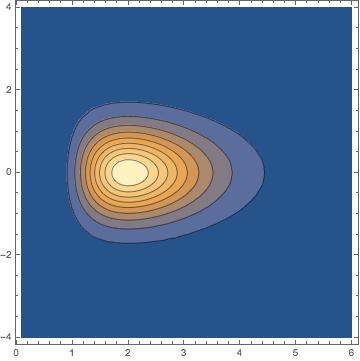} 
\includegraphics[scale=0.33]{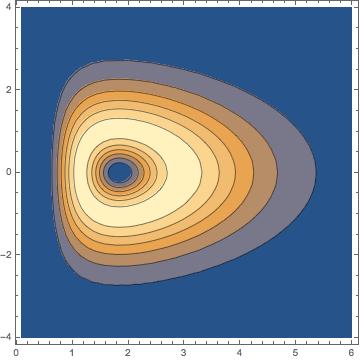}
\caption{\label{fig2} A phase space portrait  $\rho_{\psi}(q,p)$ with $\nu=3$ for $\ket{\psi} = \ket{q=2,p=0, \nu, n=0}$ (left panel) and $\ket{\psi} = \ket{q=2,p=0, \nu, n=1}$ (right panel).}
\end{center}
\end{figure}

For $n \ne 0$, the functions $x \mapsto |\scalar{x}{q,p; \nu,n}|^2$ have several maxima and zeros, so that these states are highly non-classical, although their structure is always stable with time. Fig. \ref{fig2}) shows the semi-classical probability density $\rho_{\psi}(q,p) = |\scalar{q,p; \nu,0}{\psi}|^2/(2 \pi c_0)$ for $\ket{\psi} = \ket{q=2,p=0, \nu, n=1}$ compared with the one obtained for $\ket{\psi} = \ket{q=2,p=0, \nu, n=0}$, i.e., keeping the same parameters at the exception of $n$. While in the case $n=0$ the structure is simple with a well-defined peak reaching its maximum at one point, at the opposite for $n=1$ the structure is more complex with always a peak but equipped with a central hole and a ring for maximal values.

\section{Conclusion}\label{SecConc}

Focusing on the case of a classical Hamiltonian $H=p^2$ on the half-plane, we discovered a class of dynamically stable coherent states parametrized by the half-plane phase space. They are expected to play an important role in quantum cosmology, e.g., by providing exact expressions for expectations values of compound dynamical quantum observables. The latter are requisite for coupling primordial perturbations to quantum models of the universe. 

Moreover, these new coherent states enhance the framework of ACS quantization. By fixing the choice of the fiducial vector they remove some ambiguous aspects of the framework. On one hand, the new ACS evolve parametrically through the Schr\"{o}dinger equation, with the parameters following a trajectory in phase-space given by a semi-classical Hamiltonian. On the other hand, the fiducial vector usually unspecified is now given as an eigenvector of a well-defined Hamiltonian. The basic quantum observables given by this enhanced procedure remain unchanged, up to some normalization factors. Therefore, the new ACS are really ``exceptional'' states among all possible definitions of affine coherent states. If one accepts the general framework of coherent state quantization as an admissible one, it becomes clear that these new coherent states should be favored.  The crucial point is that they are explicit solutions to the Schr\"{o}dinger equation, allowing to avoid troublesome approximations. We plan to use them in our future works on quantum cosmology.

\begin{acknowledgements}
P.M. acknowledges the support of the National Science Centre
(NCN, Poland) under the research Grant 2018/30/E/ST2/00370.
\end{acknowledgements}

\begin{appendix}
\section{Quantization formula}
\label{append1}

In this appendix, we sketch the proof for the formula of section \ref{secquantization}. To obtain the operators $\hat{A}_f$, we need to calculate the matrix elements $\bra{x} \hat{A}_f \ket{x'}$, i.e., to calculate in the sense of distributions, the integrals
$$
\bra{x} \hat{A}_f \ket{x'} = \int_{\Pi_+} \frac{\dd q \dd p}{2 \pi c_0} \scalar{x}{q,p} \scalar{q,p}{x'} f(q,p).
$$
For the cases mentioned in \ref{secquantization}, i.e. $f(q,p)= q^\alpha, p, qp, p^2$, $f(q,p)$ is factorized as a product of a function of $q$ by a function of $p$, and the technique is the same. We have
\begin{widetext}
\begin{equation}
\bra{x} \hat{A}_{q^\alpha p^\beta} \ket{x'} = \int_{\Pi_+} \frac{\dd q \dd p}{2\pi c_0} q^\alpha p^\beta \exp \left[ \frac{\ii p}{2q} (x^2-x'^2) \right] \frac{1}{q} \psi_0 \left( \frac{x}{q}\right) \psi_0\left( \frac{x'}{q}\right).
\end{equation}
First we can use a change of canonical variables $\left[ Q = q^2, P = p/(2q)\right]$, leading to
\begin{equation}
\bra{x} \hat{A}_{q^\alpha p^\beta} \ket{x'} = 2^\beta \int_{\Pi_+} \frac{\dd Q \dd P}{2\pi c_0} Q^{\frac12 \left(\alpha+\beta\right)} P^\beta \exp \left[ \ii P (x^2-x'^2) \right] \frac{1}{\sqrt{Q}} \psi_0\left( \frac{x}{\sqrt{Q}}\right) \psi_0\left( \frac{x'}{\sqrt{Q}}\right).
\end{equation}
Therefore we have
\begin{equation}
\label{quantgene}
\bra{x} \hat{A}_{q^\alpha p^\beta} \ket{x'} = \frac{2^\beta}{c_0} \left[ \int_{-\infty}^{+\infty} \frac{\dd P}{2 \pi} P^\beta e^{\ii P (x^2-x'^2)} \right] \int_0^\infty \dd Q Q^{\frac12 \left(\alpha+\beta-1\right)} \psi_0\left( \frac{x}{\sqrt{Q}}\right)
\psi_0\left( \frac{x'}{\sqrt{Q}}\right).
\end{equation}
The basic relations  
$$
\int_{-\infty}^{+\infty} \frac{\dd P}{2\pi} e^{\ii P y} = \delta(y)\quad \Longrightarrow \quad \int_{-\infty}^{+\infty} \frac{dP}{2\pi} P e^{\ii P y} = -\ii \delta'(y)\quad  \text{and} \quad \int_{-\infty}^{+\infty} \frac{\dd P}{2\pi} P^2 e^{\ii P y} = - \delta''(y)
$$
show that, to calculate expressions \eqref{quantgene}, at least for $\beta=0,1,2$, we need to simplify formula of the type $f(x,a) \delta^{(\beta)}(x^2-a^2)$ with the constraint $x,a >0$.\\

Using the well-known relations for the Dirac $\delta$-function
\begin{align*}
f(x) \delta(x-a) &= f(a) \delta(x-a);\\
f(x) \delta'(x-a) &= f(a) \delta'(x-a) - f'(a) \delta(x-a); \\
f(x) \delta''(x-a) & = f(a) \delta''(x-a) -2f'(a) \delta'(x-a) +f''(a) \delta(x-a) \,,
\end{align*}
and the fact that $\delta(x^2 - a^2) = \frac{1}{2a} \delta(x-a)$ if $x,a >0$, it is straightforward  to obtain
\begin{subequations}
\begin{align}
\label{delta1} f(x) \delta(x^2-a^2) &=\frac{f(a)}{2a} \delta(x-a), \\
\label{delta2} f(x) \delta'(x^2-a^2) &= \frac{f(a)}{4a^2} \delta'(x-a) + \left[ \frac{f(a)}{4a^3} - \frac{f'(a)}{4a^2} \right] \delta(x-a),\\
\label{delta3} f(x) \delta''(x^2-a^2) & = \frac{f(a)}{8a^3} \delta''(x-a) + \left[ \frac{3 f(a)}{8 a^4} - \frac{2 f'(a)}{8 a^3}\right] \delta'(x-a) + \left[ \frac{f''(a)}{8a^3} - \frac{3f'(a)}{8a^4} + \frac{3f(a)}{8a^5} \right] \delta(x-a),
\end{align}
\end{subequations}
\end{widetext}

Let us show how for instance that Eq.~\eqref{delta1} allows to recover $\hat{A}_{q^\alpha}$. We have
\begin{equation}
\bra{x} \hat{A}_{q^\alpha} \ket{x'} = \frac{1}{c_0} \delta(x^2-x'^2) f(x,x'),
\end{equation} 
in which we set
\begin{equation} f(x,x') = \int_0^\infty \dd Q Q^{\frac{\alpha-1}{2}} \psi_0\left( \frac{x}{\sqrt{Q}}\right)
\psi_0\left( \frac{x'}{\sqrt{Q}}\right),
\end{equation}
so that, from \eqref{delta1}
\begin{equation} 
\bra{x} \hat{A}_{q^\alpha} \ket{x'} = \frac{1}{c_0} \frac{f(x',x')}{2 x'} \delta(x-x')  \,,
\end{equation} 
and 
\begin{equation}
f(x',x') = 2 (x')^{\alpha+1} c_\alpha(\psi_0),
\end{equation}
and finally
\begin{equation}
\bra{x} \hat{A}_{q^\alpha} \ket{x'} = \frac{c_\alpha(\psi_0)}{c_0(\psi_0)} (x')^\alpha \delta(x-x'),
\end{equation}
which implies 
\begin{equation}
\hat{A}_{q^\alpha} = \frac{c_\alpha(\psi_0)}{c_0(\psi_0)} \hat{x}^\alpha \,.
\end{equation}
Similarly, we find
\begin{equation}
\bra{x} \hat{A}_{p} \ket{x'} = - \frac{2\ii}{c_0} \delta'(x^2-x'^2) f(x,x'),
\end{equation}
with 
\begin{equation}
f(x,x') = \int_0^\infty \dd Q \,\psi_0\left( \frac{x}{\sqrt{Q}}\right)
\psi_0 \left( \frac{x'}{\sqrt{Q}}\right).
\end{equation}
Using \eqref{delta2} we obtain first
\begin{widetext}
\begin{equation}
\bra{x} \hat{A}_{p} \ket{x'} = - \frac{2\ii}{c_0} \left\{ \frac{f(x',x')}{4 x'^2} \delta'(x-x') + \left[ \frac{f(x',x')}{4 x'^3} - \frac{\partial_x f(x',x')}{4 x'^2} \right] \delta(x-x') \right\}.
\end{equation}
\end{widetext}
Then a change of variable and an integration by parts (assuming $\psi_0$ to be rapidly decreasing) gives
\begin{equation}
f(x',x') = 2 x'^2 c_1(\psi_0) \quad \text{and} \quad \partial_x f(x',x') = 2 x' c_1(\psi_0).
\end{equation}
Thus, we conclude that
\begin{equation}
\bra{x} \hat{A}_{p} \ket{x'} = - \ii \frac{c_1(\psi_0)}{c_0(\psi_0)} \delta'(x-x') \implies \hat{A}_p = \frac{c_1(\psi_0)}{c_0(\psi_0)} \hat{p} \,.
\end{equation}

More complicated formulas involving $p$ or $p^2$ can be obtained in a similar fashion.

\section{Detailed relations involving SU(1,1)}
\label{append2}

The basic $\mathrm{SU}(1,1)$ self-adjoint generators $\hat{K}_{0,1,2}$ on $\mathcal{H} = L^2(\mathbb{R}_+, \ud{x})$  mentioned in Sec. \ref{rationale} read in our case 
\begin{align}
 \hat{K}_0 &= \frac{1}{2}\left( \hat{H}+\frac{\hat{x}^2}{4}\right)\,,\\
 \hat{K}_1 &= \frac{1}{2}\left[ \cos\omega \left( \hat{H} - \frac{\hat{x}^2}{4}\right)+\sin\omega\,\hat{d}\right]\,, \\
 \hat{K}_2 &= \frac{1}{2}\left[ -\sin\omega \left(\hat{H} - \frac{\hat{x}^2}{4}\right) + \cos\omega\,\hat{d}\right], \end{align}
where $\hat{H} = \hat{p}^2+ \dfrac{C}{\hat{x}^2}$ with $C>0$ and $\omega$ is a free real parameter. They verify the usual commutation rules of $\mathfrak{su}(1,1)$, i.e. 
\begin{equation}
[\hat{K}_0,\hat{K}_1]= \mathrm{i}\hat{K}_2\, , \ [\hat{K}_0,\hat{K}_2]= -\mathrm{i}\hat{K}_1\,, \ [\hat{K}_1,\hat{K}_2]= -\mathrm{i}\hat{K}_0\,. 
\end{equation}
Conversely, we have, 
\begin{align}
\label{tHK}
 \hat{H}&= K_0 + \cos\omega \hat{K}_1 -\sin\omega \hat{K}_2\,,\\
\label{x2K} \displaystyle{\frac{\hat{x}^2}{4}} &= \hat{K}_0 - \cos\omega \hat{K}_1 + \sin\omega \hat{K}_2\,,\\
\label{dK} \displaystyle{\frac{\hat{d}}{2}} &= \sin\omega \hat{K}_1 +\cos\omega \hat{K}_2\, . \end{align}
In the unit disk realisation of SU$(1,1)$ actions,  $\hat{K}_0$ is the generator of rotations while $\hat{K}_1$ and $\hat{K}_2$ are generators of hyperbolic transforms. Introducing the ladder operators 
\begin{equation}
\hat{K}_{\pm}= \hat{K}_2\mp \mathrm{i} \hat{K}_1\,,\quad [\hat{K}_+,\hat{K}_-]= -2\hat{K}_0\, , 
\end{equation}
one can also write
\begin{align}
 \hat{H}&= \hat{K}_0 -\frac{1}{2\mathrm{i}}\left(\hat{K}_+ e^{\mathrm{i}\omega} -\hat{K}_- e^{-\mathrm{i}\omega}\right)\,,\\
 \displaystyle{\frac{\hat{x}^2}{4}} &= \hat{K}_0 + \frac{1}{2\mathrm{i}}\left(\hat{K}_+ e^{\mathrm{i}\omega} -\hat{K}_- e^{-\mathrm{i}\omega}\right)\,,\\
 \hat{d} &=\hat{K}_+ e^{\mathrm{i}\omega} +\hat{K}_- e^{-\mathrm{i}\omega} \, . \end{align}
The Casimir operator is given by 
\begin{equation}
 \hat{\mathcal{Q}} = \hat{K}_1^2 +\hat{K}_2^2-\hat{K}_0^2= \frac{1}{2}\{\hat{K}_+, \hat{K}_-\}-\hat{K}_0^2 \, ,
\end{equation}
which gives, in this representation
\begin{equation}
\hat{\mathcal{Q}} = \frac{1}{4} \left(\frac{3}{4}-C \right) \mathbbm{1} \quad \text{with} \quad C >0\, .
\end{equation}
Hence, the corresponding $\mathrm{SU}(1,1)$  UIR is identified through the Bargman index $\eta > 0$ such that $\eta(\eta-1)= \frac14 (C-\frac34)$, i.e. $\eta= \frac12 \pm\sqrt{C+\frac14}$ . For $C=\frac34$, which corresponds to the lowest value allowing the operator $\hat{H}$ to have a unique self-adjoint extension, we get the element lying at the bottom of the genuine discrete series corresponding to $\eta=1$. 

Since the parameter $\omega$ introduced above is free, we choose in what follows $\omega=0$ to simplify expressions and calculations.\\

With this material at hand,  the unitary operator $\hat{V}_{q,p}$ of Eq.~\eqref{UnV} can be viewed as a combination of $\mathrm{SU}(1,1)$ displacement operators \cite{gazeaubook09} allowing to interpret our CS in terms   of Perelomov CS for SU$(1,1)$. We first have, from Eq.~\eqref{UnV}
\begin{equation}
\label{Vqpfirst}
\hat{V}_{q,p}=e^{2 \mathrm{i} p(\hat{K}_0-\hat{K}_1)/q} e^{-2 \mathrm{i} \ln q \hat{K}_2} \,.
\end{equation}
One can rewrite $\hat{V}_{q,p}$ differently in terms of the generators (see below) as
 \begin{equation}
 \label{desintangl}
  \hat{V}_{q,p} = e^{(\xi \hat{K}_+ -\bar\xi \hat{K}_-)} \,e^{\mathrm{i}\theta \hat{K}_0}=e^{\mathrm{i}\theta^{\prime} \hat{K}_0} e^{(\xi^{\prime} \hat{K}_+ -\bar{\xi^{\prime}} \hat{K}_-)}\, .   \end{equation}
The  unitary operator $e^{(\xi \hat{K}_+ -\bar\xi \hat{K}_-)}$ is the SU$(1,1)$ analogue of the 
displacement operator in the Weyl-Heisenberg symmetry case, and was used by Perelomov to build his  SU$(1,1)$ coherent states \cite{perelomov86,perelomov:1972cmp}. On the other hand, the exponentiation of the operator $\hat{K}_0$, i.e. $e^{\mathrm{i}\theta \hat{K}_0}$, leads to the compact subgroup of rotations in the unit disk. In doing so, we use the one-to-one correspondence 
\begin{subequations}
\begin{align}
    \mathrm{i} \hat{K}_0 &\leftrightarrow N_0 = \frac{1}{2}\begin{pmatrix}  \mathrm{i} & 0\\
 0& -\mathrm{i} \end{pmatrix}\, ,\\
 \mathrm{i} \hat{K}_1 &\leftrightarrow N_1 = \frac{1}{2}\begin{pmatrix}  0 & 1\\
 1& 0\end{pmatrix}\, ,\\
 \mathrm{i} \hat{K}_2 &\leftrightarrow N_2 = \frac{1}{2}\begin{pmatrix} 0 & \mathrm{i} \\
  -\mathrm{i}&0 \end{pmatrix},
\end{align}
\end{subequations}
i.e.,
\begin{equation}
  \mathrm{i} \hat{K}_+ \leftrightarrow N_+ =\begin{pmatrix} 0  & 0\\
  -\mathrm{i} & 0\end{pmatrix}\quad \text{and} \quad \mathrm{i} \hat{K}_- \leftrightarrow N_- =\begin{pmatrix} 0  & \mathrm{i}\\
  0 & 0\end{pmatrix}.
\end{equation}
 Then, one  uses the generic exponentiation correspondence between $X\in \mathfrak{su}(1,1)$ and $g\in \mathrm{SU}(1,1)$,
 \begin{align*}
 X &= \sum_{i=0,1,2}\lambda_i N_i = \frac{1}{2}\begin{pmatrix}
      \ii\,\lambda_0 & z\\
      \bar z & -\ii\,\lambda_0
  \end{pmatrix}\, , \ z=\lambda_1 +\ii \,\lambda_2\,, \\
  &\mapsto g= \exp X= \begin{pmatrix}
      \alpha &\beta\\
      \bar\beta & \bar\alpha
  \end{pmatrix}\, , \ \vert\alpha\vert^2-\vert\beta\vert^2=1, 
\end{align*}
so that, for $\hat{V}_{q,p}$ of Eq.~\eqref{Vqpfirst}, we have the correspondences between unitary operators $\mapsto$ $2\times 2$ matrices $\in \mathrm{SU}(1,1)$:
 \begin{equation}
  e^{2 \ii p (\hat{K}_0-\hat{K}_1)/q}\mapsto \begin{pmatrix}
      1+\ii \,\displaystyle\frac{p}{q}& - \displaystyle\frac{p}{q}\\ & \\
  - \displaystyle\frac{p}{q} &  1-\ii \displaystyle\frac{p}{q}   \end{pmatrix}\, ,    
 \end{equation}
 \begin{equation}
  e^{-2 \ii\ln q\,\hat{K}_2}\mapsto \begin{pmatrix}
      \displaystyle\frac{q + q^{-1}}{2}& -\ii \, \displaystyle\frac{q - q^{-1}}{2}\\ & \\
\ii\, \displaystyle\frac{q - q^{-1}}{2}   &  \displaystyle\frac{q + q^{-1}}{2}   \end{pmatrix},   
 \end{equation}
 and so
 \begin{equation}
  \hat{V}_{q,p}\mapsto \begin{pmatrix}
      \alpha & \beta\\
      \bar\beta & \bar\alpha
  \end{pmatrix},
\end{equation}
 with
 \begin{equation}
  \alpha = \frac{q + q^{-1}}{2} + \ii \,\frac{p}{q^2} \quad \text{and} \quad \beta= -\ii\, \left( \frac{q - q^{-1}}{2} -\ii\,\frac{p}{q^2} \right)\,. 
 \end{equation}
 We also have the left and right Cartan factorizations of SU$(1,1)$, namely
 \begin{align*}
  \begin{pmatrix}
      \alpha & \beta\\
      \bar\beta & \bar\alpha
  \end{pmatrix}&= \begin{pmatrix}
      \delta & \delta\zeta\\
      \delta\bar\zeta & \delta
  \end{pmatrix}\begin{pmatrix}
      e^{\ii\frac{\theta}{2}} &0\\
      0 & e^{-\ii\frac{\theta}{2}}
  \end{pmatrix}\equiv p(\zeta)\,h(\theta)\\ 
  &= \begin{pmatrix}
      e^{-\ii\frac{\theta}{2}} &0\\
      0 & e^{\ii\frac{\theta}{2}}\end{pmatrix} \begin{pmatrix}
      \delta & \delta\zeta^\prime\\
      \delta\bar\zeta^\prime & \delta
  \end{pmatrix}\equiv h(-\theta)\,p(\zeta^\prime)\, ,  
 \end{align*}
 with $e^{\ii\frac{\theta}{2}}=\frac{\alpha}{\vert\alpha\vert}$, $\zeta=  \beta\bar\alpha^{-1}$, $\zeta^\prime=  \beta\alpha^{-1}$, and $\delta=\vert\alpha\vert=(1-\vert\zeta\vert^2)^{-1/2}$ .
 In terms of exponentials of $\mathfrak{su}(1,1)$ elements and corresponding unitary operators, the decomposition factors read
 \begin{align*}
  h(\theta)&=  e^{\theta N_0}\mapsto e^{-\ii \theta K_0}\, , \\
  p(\zeta) &= e^{-\ii\xi N_+ + \ii\bar\zeta N_-}\mapsto e^{\xi K_+ - \bar\xi K_-}\, , 
 \end{align*}
with $\zeta= -\tanh\vert\xi\vert\,\exp(-\ii \arg\xi)$. The application of these relations to \eqref{desintangl} yields quite involved expressions in terms of the original variables $q$ and $p$ whose explicit form we do not write here. 
 
\end{appendix}

\bibliography{references}
\end{document}